\documentstyle[11pt]{article}
\newcommand{\newsubsection}[1]{
\vspace{1cm}
\pagebreak[3]
\addtocounter{subsection}{1}
\addcontentsline{toc}{subsection}{\protect
\numberline{\arabic{subsection}}{#1}}
\noindent{ \sc #1}                  
\nopagebreak
\vspace{2mm}
\nopagebreak}

\newlength{\extraspace}
\newlength{\extraspaces}
\newcommand{\be}{\begin{equation}
\addtolength{\abovedisplayskip}{\extraspaces}
\addtolength{\belowdisplayskip}{\extraspaces}
\addtolength{\abovedisplayshortskip}{\extraspace}
\addtolength{\belowdisplayshortskip}{\extraspace}}
\newcommand{\ee}{\end{equation}}
\newcommand{\ba}{\begin{eqnarray}
\addtolength{\abovedisplayskip}{\extraspaces}
\addtolength{\belowdisplayskip}{\extraspaces}
\addtolength{\abovedisplayshortskip}{\extraspace}
\addtolength{\belowdisplayshortskip}{\extraspace}}
\newcommand{\ea}{\end{eqnarray}}
\newcommand{\nonu}{\nonumber \\[.5mm]}
\newcommand{\is}{& \!\! = \!\! &}
\newcommand{\Bl}{\Bigl\langle}
\newcommand{\Br}{\Bigr\rangle}
\renewcommand{\d}{{{\partial}}}

\newcommand{\half}{{\textstyle{1\over 2}}}

\newcommand{\ra}{\rightarrow}

\newcommand{\tr}{{\rm tr}}
\newcommand{\sss}{{\mbox{\footnotesize  S}}}

\newcommand{\ddd}{\dot}
\newcommand{\dddd}{D}%{{\mbox{\footnotesize  D}}}
\newcommand{\xxx}{{\mbox{\footnotesize  X}}}

\input epsf
\begin{document}
\begin{flushright}
{\small April 1997}\\
{\small UTFA-97/16}\\[2mm]
\end{flushright}
\begin{center}
{\Large\sc {A Matrix String Interpretation\\[2.5mm]
             of the Large N Loop Equation}}\\[12mm]
{\large  Herman Verlinde}\\[3mm]
{\it Institute for Theoretical Physics\\[2mm]
 University of Amsterdam, 1018 XE Amsterdam} \\[11mm]
{\bf Abstract}
\end{center}
\noindent
The existence two S-dual descriptions of (N,1) string bound states 
suggests that the strong coupling behaviour of electric flux lines in large 
N 2D SYM theory has a dual description in terms of weakly coupled IIB 
string theory. In support of this identification, we propose a dual 
interpretation of the SYM loop equation as a perturbative string Ward 
identity, expressing the conformal invariance of the corresponding 
boundary interaction. This correspondence can be viewed as a weak 
coupling check of the matrix string conjecture.

\newcommand{\gym}{ g_{\!{}_{Y\! M}}}
\medskip

\newsubsection{Introduction}

Consider the large N limit of 1+1-dimensional ${\cal N} \! =\!  8$
supersymmetric Yang-Mills theory with gauge group U(N), described by the Lagrangian
(here and in most of this paper we omit all fermionic fields)
\be
\label{sym}
S = \int \! d^2\xxx \; \tr\Bigl[ \
{1\over \gym^2} F_{\alpha\beta}^2 + (D_\alpha \Phi^i)^2 - 
{\gym^2}[\Phi^i,\Phi^j]^2 
\Bigr].
%+ \theta^T \gamma^\alpha
%D_\alpha \theta + \gym \theta^T\gamma_i [\Phi^i,\theta]\right].
\ee
with $i = 2, \ldots, 9$. This model can be viewed as the 
dimensional reduction of ten dimensional ${\cal N}\! =\! 1$ SYM theory,
where $\Phi^i$ represent the 8 transverse components of the 
ten-dimensional gauge potential $A_\mu$.
We will be interested in giving a string interpretation
of the loop equation of motion \cite{polyakov,migdal} 
of the Wilson loop average 
\be
\label{loop}
W[\xxx] = \Bl \tr P \exp \oint_C\!\! 
d\sss\, \ddd \xxx^\mu A_\mu(\xxx(\sss)) \Br
\ee
in this 2D SYM theory. Here 
$\sss = (s,\theta)$ denotes the 1D super-space coordinate
along the loop $C$. % and $\ddd= \d_\theta + \theta \d_s$.
The index $\mu$ in (\ref{loop}) runs from $0$ to $9$,
so that the path-ordered exponential 
is in fact defined along a path in ten dimensions, where the ${\cal N}\!=\! 
1$ $D\! =\! 10$ gauge field $A_\mu  = (A_\alpha, \Phi^i)$ 
is restricted to depend on $\xxx_0$ and $\xxx_1$ only. We will
assume that the $\xxx^1$-direction is compact, so that we can choose 
$\xxx_1$ to be periodic modulo $2\pi R$.

The above SYM model arises in ten-dimensional IIB string theory
as the zero slope $\alpha' \rightarrow 0$ description of a
bound state consisting of N D-strings. As emphasized in \cite{witten}, the 
invariance under the gauge transformations 
$B \rightarrow B + d\Lambda$ and $A \rightarrow A+ \Lambda$ of the NS 
anti-symmetric $B$-field implies that in this context the
Wilson loop (\ref{loop}) is necessarily accompanied by a IIB string world-sheet 
that fills up its interior. The path-ordered exponential in (\ref{loop}) 
indeed defines a reparametrization invariant functional of the 
trajectory $\{\xxx(\sss)\}$, and provides a 
conformally invariant boundary interaction for the IIB string, provided 
the gauge field $A$ satisfies the appropriate equation of motion.
To leading order in $\alpha'$ this is just the super Yang-Mills equation. 

The location of the string boundary divides the D-string world sheet into 
a vacuum outside region and an inside region with non-zero electric flux. 
As shown in \cite{witten}, the gauge sector carrying one fundamental
unit of electric flux $\int \! \tr\, E_1 =1$ can indeed be thought of as a 
bound state where one single type IIB string has been embedded into the $N$ 
D-string world sheet. In the limit of large N (and fixed string coupling), 
the energy per unit length of the electric flux is much smaller than the 
fundamental string tension. This reduction of tension was interpreted in 
\cite{witten} as a manifestation of the binding force between the two 
types of strings. Intuitively what happens is that, due to higher loop 
interactions, the IIB worldsheet inside the Wilson line developes holes, and 
starts to look like a Feynman graph of the SYM field theory. 
In fact, it is possible to take a combined large N  and zero slope limit, 
while keeping the effective string tension of the electric flux finite, 
and in this limit the identification of the multi-holed string worldsheet 
with a multi-loop Feynman graph becomes exact.

Given the recent progress in string duality, 
it seems natural to look for a dual formulation of the theory in which 
the electric flux string can be studied in a weak coupling language. 
There is by now a large body of evidence that 
IIB string theory possesses an exact duality symmetry that 
interchanges the roles of the D-strings and the fundamental strings. 
We can apply this symmetry to our context, and reinterpret the 
two-dimensional SYM model as a ``matrix string'' description of a 
collection of N fundamental IIB strings stretched in the 
$\xxx_1$-direction.\footnote{For notational simplicity we will assume 
that the target space coordinate $\xxx^1$ is identified with the
worldsheet coordinate $\xxx^1$. Hence the radius $R$ in fact denotes
the compactification radius of the target space $S^1$.}
In this new identification, the eigenvalues of the Higgs
fields $\Phi^i$ parametrize the location of the fundamental IIB strings, 
while the SYM electric flux becomes identified with the D-string charge.
Unless stated otherwise, we will from now on apply this switch in terminology!

The Yang-Mills coupling $\gym$ is expressed in terms of the IIB string 
coupling and slope parameter as 
\be
{1\over \gym^{2}} = {g}_s^2 {\alpha}' 
\ee
A direct way to obtain this relation is by equating the tension of a 
single unit of electric flux with the effective tension of a single D-string 
embedded inside N IIB strings. The tension of the combined (N,1) string 
bound state reads
\be
\label{tensionb}
T \; = \ {1\over {\alpha}'} \; \sqrt{N^2 + {1\over {g}_s^2}}
\ \simeq  \ {N\over {\alpha}'}  \; + \; {1\over 2 Ng_s^2 \alpha'}
\ee
The second term must be compared with the energy per unit length of one unit of
electric flux, given by $E/R = \gym^2 /2N$. We note that, similarly 
as before, the embedded D-string has reduced its string tension 
to $T = 1/\alpha_{ef\! f}$ with
\be
\label{aleph}
\alpha_{ef\! f} = 2  N \, g_s^2 \, \alpha'
\ee
which amounts to a renormalization by a factor of $2Ng_s$ relative to the 
tension $T = (g_s \alpha')^{-1}$ of a free D-string. In the following we
will combine the large N limit with a rescaling of $g_s$ and $\alpha'$,
such that the effective string scale (\ref{aleph}) of the electric flux 
string remains finite.

We can compare this dual representation of the (N,1) string with 
the conventional description in terms perturbative IIB string theory 
\cite{witten}. From the latter viewpoint, the worldsheet dynamics of 
the D-string is induced by integrating out the open IIB strings attached 
to it. The general boundary interaction reads 
\be
\label{bound}
\oint d\sigma \Bigl[\; \tilde{A}_\alpha(\xxx) \partial_\tau \xxx^\alpha 
+ Y_i(\xxx) \partial_\sigma \xxx^i\Bigr]
\ee
with $\alpha = 0,1$. The fields ${Y}_i(\xxx)$ parametrize the transverse 
motion of the D-string, while $\tilde{A}_1(\xxx)$ specifies an abelian 
gauge field defined on its worldsheet.
Standard perturbative methods \cite{borninfeld} show that their effective 
dynamics is 
described by the BI lagrangian
\be
\label{bi}
{\cal L}_{BI} = {1\over {g}_s\alpha'} 
\sqrt{\det(G_{\alpha\beta}+
\alpha' \tilde{F}_{\alpha\beta})}
\ee
where 
$G_{\alpha\beta} = 
\delta_{\alpha\beta}+ \partial_\alpha Y^i \partial_\beta Y_i$ denotes
the induced metric on the D-string worldsheet.

Working in the $\tilde{A}_0\!=\!0$ gauge, we have 
$\tilde{F}_{01} = \partial_0\tilde{A_1}$.
Due to the presence of large gauge transformations, the potential $\tilde{A}_1$ 
defines a periodic variable. The abelian electric field $\tilde{E}_1$, defined as 
the canonical momentum conjugate to $\tilde{A}_1$
\be
\tilde{E}_1 = {\tilde{F}_{01} 
\over g_s \sqrt{\det(G+ \alpha'\tilde{F})}},
\ee 
therefore has quantized total flux
\be
\label{fluxs}
{1\over R} \oint \! dx_1\, \tilde{E}_1 = N 
\ee
which 
according to \cite{witten} needs to be interpreted as the number of IIB 
strings bound to the D-string. The tension (\ref{tensionb})
of the total bound state is equal to the Hamiltonian density 
${\cal H} = E \dot{A_1} - {\cal L}_{BI}$ evaluated in the ground state 
of the flux sector (\ref{fluxs}), see e.g. \cite{curtigor}.

It will be useful to visualize this dual representation of the large N 
limit.
Taking N to infinity in this case corresponds to the limit
in which the BI field strength $\tilde{F}_{01} = \partial_0\tilde{A}_1$ 
approaches its {\it critical} value $F_{01}=1/\alpha'$. 
The physical meaning of this critical 
limit is that the end points of the open IIB strings, which are 
electrically charged, are pulled very far away from each other  under influence 
of the critical field. Hence the IIB worldsheet is stretched to a very 
large size compared to the original string scale set by $\alpha'$. 
Mathematically, the boundary interaction (\ref{bound}) in a constant
electric field gives rise to a term $F_{01} \oint \xxx^0\partial 
\xxx^1$, which equals $F_{01}$ times the 2D area covered 
by the IIB string world sheet. In the critical limit, this contribution 
almost cancels the standard Nambu-Goto area. As a result, the IIB string 
tension is effectively reduced by a factor $1\!-\alpha' F_{01}$, leading 
to a new effective tension\footnote{I thank I. Klebanov and
A. Polyakov for valuable discussions on this point.} 
\be
\label{teff}
T_{\rm eff} \simeq {1\over N^2 g_s^2 \alpha'}
\ee 
The corresponding new string scale is very large, even relative to the 
scale (\ref{aleph}) of the SYM electric flux string.

We note that the renormalization of the string scale (\ref{aleph}) 
are with the same factor $2Ng_s$. In fact, an identical factor also appears 
if one computes the disk vacuum amplitude in the near critical electric field. 
One finds
\be
\label{disk}
\Bl \; 1 \; \Br_\Sigma \; \simeq \int\! d^2 x \; {\cal L}_{BI} \; \simeq 
\;  {1\over 2 Ng_s^2 \alpha'} \; \int \! d^2 x.
\ee
This should be compared with the standard normalization of the open string 
vacuum amplitude without electric field, which is $1/g_s \int \! d^2 x$. 
In (\ref{disk}) we introduced the notation $\bigl\langle \; \ldots \; 
\bigr\rangle_\Sigma$ for the
tree-level expectation value of the open IIB string moving in the 
electric field $E_1=N$.
The result (\ref{disk}) will be of relevance later on.\footnote{It is
suggestive that the result (\ref{disk}) can be written as $T_{\rm eff}
\over g_s^{\rm eff}$ with $g^{\rm eff}_s = 1/N$ and $T_{\rm eff}$ as
in (\ref{teff}) \cite{Gukov}. 
Notice further that this single string effective tension
$T_{\rm eff}$ is smaller by a factor of (order) $N$ than that of the 
electric flux string, which suggests that the electric flux string is
made up from (of order) $N$ dual open strings.}

\newsubsection{Loop equation as conformal Ward identity}

A priori, the U(N) SYM representation of the (N,1) string can be
motivated only as a dual, effective description at strong coupling. 
According to the matrix string conjecture \cite{banks,matrix-string}, 
however, this description is supposed to be valid all the way into the 
perturbative string regime (provided we send N to infinity). Though quite
convincing, the quantitative evidence supporting the equivalence of 
these two dual descriptions is still rather limited, however.
Quite generally, perturbative IIB string theory is expected to 
arise as an effective description of the infra-red SYM dynamics.
As shown in \cite{matrix-string}, in the leading IR limit the SYM 
degrees of freedom indeed reduce to that of a second quantized 
gas of freely propagating strings\footnote{The discussion in \cite{matrix-string}
concerned the matrix representation \cite{banks,motl} of type IIA strings,
which is related via T-duality along the $\xxx^1$-direction to the present 
IIB set-up. We will briefly discuss this T-duality map in the concluding 
section.} In addition, evidence was presented
that the perturbative splitting and joining interactions between the 
strings is also reproduced. We would like to find a similar 
correspondence in the present context. 
In particular, we would like to see how the gauge theory equation of motion
is reproduced in the dual language. These equations of motion are 
expressed in terms of the loop average (\ref{loop}) via the well-known 
loop equation \cite{polyakov,migdal}. A short summary is given in Appendix A.

Since the SYM electric flux translates into D-string 
winding under the duality, the Wilson loop now marks the boundary of 
a piece of open D-string attached to the N fundamental strings. In the IR
regime, the IIB strings formed by the Higgs eigenvalues should therefore
contain sector of open strings, whose boundaries are confined to the 
interior of the Wilson loop. Although it should be relatively 
straightforward to exhibit this open string sector in the SYM language 
(along the lines of \cite{matrix-string}), we will at this point simply 
assume that this sector exists and that its properties are exactly
as described above in terms of strings in a near-critical U(1) electric 
field.

We will assume that the fundamental string scale (set by $\alpha'$) is much  
smaller than the size of the Wilson loop (set by $\alpha_{ef\! f}$ 
in (\ref{aleph})). Nonetheless, the critical electric field stretches 
the perturbative open string worldsheet until it essentially fills the 
whole interior region of the loop $C$. It is therefore reasonable to 
expect that the interaction between the Wilson loop and the open 
IIB strings translates into an effective boundary condition,  
that in the IR limit will flow to a conformally invariant fixed point. 
In the following we will investigate the associated string equations of 
motion, and exhibit a detailed correspondence with the large N SYM loop 
equation.

Instead of the position representation (\ref{loop}) of the Wilson loop,
it will turn out to be convenient to introduce a generalized 
fourier transform of the loop average, formally defined as 
({\it cf.} \cite{migdal})
\be 
\label{transform}
W_n(\epsilon_i,k_i) = %^{\mu_1, \ldots, \mu_n}(k_1, \ldots , k_n) =
\int [d\xxx]  \, \; W[\xxx] \; \, \prod_{i=1}
V_{\epsilon_i}(k_i)
%\int_{s_1}^1 \! d\sss_2 V^{\mu_2}(k_2,\sss_2) \ldots
%\int_{s_{n-1}}^1 \! d\sss_n V^{\mu_n}(k_n,\sss_n)
\ee
Here
\be
V_{\epsilon_i}(k_i) = \oint_C \! d\sss_i \, V_{\epsilon_i}(k_i,\sss_i)
\ee
with
\be
V_{\epsilon_i}(k_i,\sss_i) =
 \epsilon_\mu \ddd \xxx^\mu(\sss_i) e^{ik_i \xxx(\sss_i)}. 
\label{vertex}
\ee
The integrand on the right-hand side is reparametrization 
invariant, and the functional integration over the paths $\xxx(\sss)$ 
needs to be gauge fixed accordingly. The SYM loop equation of the position 
loop $W[\xxx]$ implies a recursion relation for the transformed amplitudes 
$W_n(\epsilon_i, k_i)$, which we have written in equation (\ref{newloop}) in 
Appendix A. 
The form of this recursion relation (\ref{newloop})
is very suggestive of a non-linear Ward identity expressing the 
cut-off independence of a collection of string amplitudes. In the following
we will summarize the reasoning that supports this interpretation.

To start with, we imagine that the transformed Wilson averages (\ref{transform}) 
can be given a dual representation in the perturbative IIB string theory by means 
of an analogous expectation value
\be 
\label{amplitude}
W_n(\epsilon_i,k_i) = \Bl \; \prod_{j} 
V_{\epsilon_j}(k_j) \; \Br_\Sigma
\ee
of local photon vertex operators (\ref{vertex}). Usually one only considers 
string amplitudes of vertex operators $V_\epsilon(k)$  that
create on-shell asymptotic states. 
The transformed loop average $W_n(\epsilon_i, k_i)$ defined above, however, 
is clearly an off-shell quantity. It will therefore be crucial 
for our set-up that the vertex operators $V_\epsilon(k)$ are 
not exactly on-shell. Nonetheless, as we will argue, it is possible
to impose the condition of conformal invariance on the amplitudes and 
find non-trivial solutions. Under an infinitesimal reparametrization 
$\sss \ra \sss+\xi(\sss)$ we have
\be
\delta_\xi \Bl \prod_{j} 
V_{\epsilon_j}(k_j) \ \Br_\Sigma = \sum_i  \Bl \; \prod_{j\neq i} 
V_{\epsilon_j}(k_j) \ 
\Bigl[L[\xi],   V_{\epsilon_i}(k_i)\Bigr]\;  %\prod_{i< j} V_{\epsilon_j}(k_j) 
\Br_\Sigma 
\label{linear}
\ee
where
\be 
L[\xi] = \oint \! d\sss\; \xi^{| |}(\sss)\, 
T_{\perp |  | }(\sss)
\ee
is a tangential component of the world-sheet stress-energy tensor. 
The conformal transformation law of $V_\mu(k,\sss)$
reads
\ba
\label{variation}
\Bigl[\, L[\xi] , V_\mu(k,\sss) \, \Bigr]  
\is
\half  \;
(k^2\delta_{\mu\nu} \!\! - \! k_\mu k_\nu)\, \dot\xi(\sss)   \,  V_{\nu}(k;\sss) \nonu 
& & \qquad \qquad \ 
 +  \, \dddd\Bigl(\dot\xi(\sss) \, k_\mu V_T(k;\sss)
\; + \; \xi(\sss) V_\mu(k,\sss) \Bigr)
\nonumber 
\ea
with $\dddd= \d_\theta + \theta \d_s$ and
\be
\label{tachyon}
V_T(k,\sss) = e^{ik X(\sss)}.
\ee 
The first two terms of the expression on the right-hand side of 
(\ref{variation}) can be re-written as
\be
\label{variation2}
-\half \oint d\sss' \; \dot\xi(\sss') \; {\cal L}(\sss')\;   V_\mu(k,\sss)
\ee
where ${\cal L}(\sss)$ 
denotes the operator (\ref{alice}) that appears in the large N loop equation
(see also eqn (\ref{var}) in Appendix A).
\be
\label{alice2}
{\cal L}(\sss)= \int_{\strut \! -\epsilon}^{\epsilon}\!\! 
d \sss' \, {{\delta^2} \over{\delta \xxx_\mu(\sss+\sss')\delta  
\xxx_\mu(\sss)}} .
\ee 
This correspondence is a first indication that the loop 
equation can be interpreted as a condition of conformal invariance.

As seen in equations (\ref{variation}), 
the insertion of the stress-tensor 
gives rise to two types  of contributions: besides the anomalous conformal 
dimension of the vertex operators, we distinguish a total derivative term. 
Very naively, one could drop this total derivative term and arrive at
the usual linear on-shell condition for the vertex operators as the condition
for conformal invariance. String theory, however, is a non-linear theory, and 
the linear on-shell condition can receive corrections from string interactions. 
These corrections arise from the total derivative term, which
can give contributions from the boundary of the $\sss_i$ 
integration domain (that is, when two or more vertex operators collide). 
These additional factorization terms 
will turn the Ward identity into an non-linear recursion relation.

The moduli space ${\cal M}_{D,n}$
of the disk $D$ with $n$ boundary points is not simply equal to
the product of $n$ copies of the disk boundary $\partial D$.
In particular, it has a boundary $\partial {\cal M}_{D,n}$, which
is reached when points start to collide. Naively one would 
think this boundary parametrizes configurations
with only two points colliding, since collisions of
more than two points appear to be of higher co-dimension.
However, a more natural compactification of ${\cal M}_{D,n}$
is obtained by using the full conformal group of the
complex plane to transform the collision between two or more boundary
points on the disk into a conformally equivalent situation, in which
the single disk $D$ is about to split into two smaller disks $D_1$
and $D_2$. There is one such boundary component for each way of
dividing the boundary points into two smaller groups.
Thus, schematically
\be
\partial {\cal M}_{D,n} =
\sum_{m=2}^{n-2} {\cal M}_{D,m+1} \times {\cal M}_{D,n-m+1}
\ee

Due to this form of the moduli space, it not entirely correct to
represent the integral over the positions $\sss_i$ of the photon vertex
operators as contour integrations over the disk boundary. A
more correct definition of the amplitude, that extends 
also to the boundary of ${\cal M}_{D,n}$, is given in Appendix B.
There we also outline the BRST derivation of the identity that 
expresses the decoupling of the stress-energy tensor from physical 
correlators.

In this Ward identity, at each given component of $\partial 
{\cal M}_{D,n}$, one finds a factorization term whenever 
a physical state propagates through the small strip separating 
the two disks. Since the momenta $k_i$
that we consider are assumed small 
compared to the fundamental string scale set by $\alpha'$, only the 
massless photon states can become on-shell. (All excited string states
still have masses of the order of $1/\sqrt{\alpha'}$.) Combining all 
possible terms, we indeed arrive at a non-linear Ward identity of the 
same form as the equation (\ref{newloop}) derived from the 
loop equation. 
\ba
\label{new}
& & \half \sum_{j=1}^n \epsilon_j^{\, \mu} 
(k^2\delta_{\mu\nu} - k_\mu k_\nu)
\Bigl\langle V^\nu(k_j, 0)
\prod_{i\neq j} V_{\epsilon_i}(k_i) \Bigr\rangle_\Sigma
=   \\[1.5mm]
%\atop {\cal J}_1 \sqcup{\cal J}_2 ={\cal J}}}
& & \qquad \qquad  
\sum_{{\cal I} \sqcup {\cal J} = \{1,\ldots,n\}}
\Bigl\langle \prod_{i\in {\cal I}} V_{\epsilon_i}(k_i)
\,V_\nu(q_{\cal I}, 0)\Bigr\rangle_\Sigma
\Bigl\langle \,V_\nu(-q_{\cal I})
\prod_{j\in {\cal J}} V_{\epsilon_j}(k_j)\Bigr\rangle_\Sigma
\nonumber
\ea
with $q = - \sum_{i\in {\cal I}} k_i$.

Another way of obtaining this result is to consider the 
behaviour of the amplitudes (\ref{amplitude}) under conformal 
transformations near the disk boundary.  
In general, the expectation value must be regulated 
and this introduces a dependence on a short-distance cut-off.
If we define this cut-off in terms of some fixed small coordinate
difference, this results in a dependence on the local coordinate
$\sss$ that parametrizes the boundary $C$. One source of this 
coordinate dependence is the anomalous scale dimension of the
vertex operators $V_{\epsilon_i}(k_i)$. Another cut-off dependence
arises because, due to the singular OPE's between the boundary vertex 
operators, the integral over the $\sss_i$ must be regulated 
near the boundary $\partial{\cal M}$. We can again specify this 
cut-off in terms of a fixed small coordinate distance.
As a result, the amplitude receives an additional non-linear coordinate 
dependence that should cancel the linear dependence due to the anomalous
scale dimension of the vertex operators. The non-linear
recursion equation that expresses this cancelation again 
takes the form as given in (\ref{new}).

As a final important technical comment, we note that the result 
(\ref{disk}) for the normalization of the disk vacuum amplitude 
ensures that the relative normalization of the left and right-hand 
side of this non-linear Ward identity (\ref{new}) is in accordance with 
that of the transformed loop-equation (\ref{newloop})
given in Appendix A.

\newsubsection{Concluding remarks}

We have used the existence two S-dual descriptions of (N,1) string bound 
states to relate the strong coupling dynamics of electric flux lines in large 
N 2D SYM theory to that of an open piece of D-string in weakly coupled IIB 
string theory. In support of this identification, we have shown that the 
SYM loop equation can be written in the form of a string Ward identity 
expressing the cut-off (in)dependence of the corresponding boundary interaction. 
Such a relation between a string Ward identity and the YM loop equation 
has been suggested on many occasions by A. Polyakov.
Our recursion relation is indeed very similar to the non-linear
$\beta$-function equations discussed in \cite{polyakov2}.
Our results indicate that the particular
example of 2D SYM theory may provide a concrete realization of these 
ideas. A crucial ingredient seems to be that
the dual IIB string propagates in a near critical electric field. This
fact allows one to consider string worldsheets large relative to the
fundamental string scale. Clearly, however, many aspects of this long distance
string correspondence with large N SYM theory still need to be clarified.

Finally, let us comment on how this result may be viewed as
a possible check of the matrix string formalism of IIA string theory
\cite{banks,motl,matrix-string}.
To arrive in the IIA context, we just need to apply a T-duality
transformation along the $\xxx^1$ direction. Instead of IIB
strings in a critical electric field, the perturbative 
description then becomes that of open IIA strings attached to
a boosted D-particle with light-cone momentum proportional to $N$. 
Furthermore, the $k_1$-components of the momenta $k_i$ in the 
transformed loop averages $W_n(\epsilon_i,k_i)$ get reinterpreted as 
shifts of the $\xxx_1$-{\it position} of this D-particle. 
In other words, the boundary of the open IIA string worldsheet
now forms a closed contour in the ($\xxx_1$, $k_0$)-plane, making
discrete jumps at the location of the vertex operators.
The analysis of the non-linear conformal Ward-identity of these
amplitudes should proceed analogously as described above, provided 
we again restrict to energies $k_0$ small compared to the string 
scale $\alpha'$ (so that only photon states can contribute in the 
factorization terms). The formal correspondence of this Ward 
identity with the SYM loop equation appears to provide new support 
for the matrix string conjecture.

\bigskip

\vspace{8mm}

{\noindent \bf Acknowledgements}

It is a pleasure to thank D. Gross, A. Mikhailov, A. Migdal, R. Dijkgraaf, 
I. Klebanov, S. Ramgoolam, E. Verlinde and especially A. Polyakov 
for very useful discussions.  This research is partly supported by a Pionier
Fellowship of NWO, a Fellowship of the Royal Dutch Academy of Sciences
(K.N.A.W.), the Packard Foundation and the A.P. Sloan Foundation.

\bigskip

\newsubsection{Appendix A: The loop equation in large N SYM theory}

Consider the path-ordered exponential 
\be
\label{loop2}
W[\xxx] = \tr P \exp \oint_C\!\! 
d\sss\, \ddd \xxx^\mu A_\mu(\xxx(\sss)) 
\ee
Here $A_\mu = (A_\alpha, \Phi^i)$ denotes the dimensional reduction
to two dimensions of the U(N) gauge potential of ten-dimensional 
${\cal N}\! =\! 1$ SYM theory. It is well-known that $W[\xxx]$ satisfies 
the following identity
\be
\label{classical}
{\cal L}(\sss) W[\xxx] =  \tr \Bigl(D^\mu F_{\mu\nu} (\xxx(\sss)) \;  
\ddd \xxx^\nu(\sss)  \;
P \exp \oint_{\strut \! C_\sss}\!\!\! d\sss'\, \ddd 
\xxx^\mu A_\mu[\xxx(\sss')]\Bigr) 
\ee
where ${\cal L}(\sss)$ denotes the functional differential operator
\be
\label{alice}
{\cal L}(\sss)= \int_{\strut \! -\epsilon}^{\epsilon}\!\! 
d \sss' \, {{\delta^2} \over{\delta \xxx_\mu(\sss+\sss')\delta  
\xxx_\mu(\sss)}} .
\ee
So if the gauge potential $A$ satisfies its Yang-Mills equation of 
motion, the classical Wilson loop satisfies ${\cal L}(\sss) W[\xxx] = 0$.
If we replace the classical Wilson line (\ref{loop}) by 
the corresponding quantum mechanical operator 
%\be
%\widehat{W}[\xxx] = \tr(P \exp \oint \ddd \xxx^\mu \widehat{A}_\mu[\xxx(\sss)])
%\Bigr\rangle_{YM}
%\ee
%where $\widehat{A}_\mu$ denotes the quantum variable of the large N SYM theory. 
the SYM-equation of motion $D^\mu F_{\mu\nu} = 0$ 
receives a quantum modification due to the contact terms
\be
(%\widehat
{D}^\mu \! %\widehat
{F}_{\mu\nu})^{a} (x) \; \, 
%\widehat
{A}^{b}_\lambda(y)\; = 
\gym^2 \delta^{ab} \delta_{\nu\lambda} \delta(x\!-\! y).
\ee
and the above linear equation for $W[\xxx]$ gets replaced by a non-linear 
identity for the Wilson loop average. 
This is the famous loop equation for the quantum mechanical 
Wilson loop \cite{migdal,polyakov}. 
In leading order at large N it factorizes into a 
closed non-linear recursion relation 
\be
\label{nloop}
{\cal L}(0)\;  %\widehat
{W}[\xxx] \, = 
{\gym^2\over N}  \int_0^1\!\!\! d \sss\, 
\ddd \xxx_\mu(0) \, \ddd \xxx_\mu(\sss) \,
\delta(\xxx(\sss)\!\! - \!\! \xxx(0)) \; 
{W}[\xxx]_0^\sss\; \;  %\widehat
{W}[\xxx]_\sss^1  \; 
\ee
where $W[\xxx]$ now denotes the SYM expectation value of (\ref{loop}).
A loop equation of this form can be derived for general 
YM theories. Although strictly speaking it is only defined in a 
regulated theory,  the high degree of supersymmetry 
in our case may be enough to ensure that this unrenormalized form 
of the equation remains valid in the continuum theory. In particular,
it is known that, by expanding $W[X]$ in powers of the gauge coupling, 
one can recover from (\ref{nloop}) the standard large N SYM perturbation
series.
In two dimensions one finds the leading order perturbative solution
\be
{W}[\xxx] \;
\simeq \exp \Bigl[{\gym^2\over 2 N} \oint \! d \sss\,\oint \! d\sss' 
\ddd \xxx_\mu(\sss) \, \ddd \xxx_\mu(\sss') \,\log |\xxx(\sss) - \xxx(\sss')| \;
\Bigr],
\ee
which shows the characteristic area law behaviour of a string with
tension $T = {\gym^2/ 2 N}$.

It is straightforward to rewrite the loop equation (\ref{loop}) 
as a recursion relation for the transformed loop averages $W_n(k_i)$
defined in (\ref{transform}). For this we need the identity
\be
\label{var}
{\cal L}(0) V_\mu(k,\sss)
=  - \delta(\sss)
(k^2\delta_{\mu\nu} - k_\mu k_\nu)  \,  V_{\nu}(k,0) 
 \; +\;  2\, D\Bigl[   
\delta(\sss) ik_\mu e^{ikX(\sss)}\Bigr]
\ee
with $D=\partial_\theta + \theta \partial_s$ and where 
the $V_\nu(k,\sss)$ the photon vertex operator defined in (\ref{vertex}).
The second, total derivative term in (\ref{var}) will 
drop out of the loop equation after the integration
over $\sss$. Using this result (and the fact that the delta-function
on the right-hand side of (\ref{nloop}) can be expanded in terms of the
photon vertices) one finds that the loop equation can be rewritten 
in the form of a closed, non-linear recursion relation of the transformed 
loop averages
\ba
\label{newloop}
& & \sum_{j=1}^n \epsilon_j^{\, \mu} 
(k^2\delta_{\mu\nu} - k_\mu k_\nu)
\Bigl\langle V^\nu(k_j, 0)
\prod_{i\neq j} V_{\epsilon_i}(k_i) \Bigr\rangle_W
=   \\[1.5mm]
%\atop {\cal J}_1 \sqcup{\cal J}_2 ={\cal J}}}
& & \qquad \qquad  {\gym^2 \over 2N}
\sum_{{\cal I} \sqcup {\cal J} = \{1,\ldots,n\}}
\Bigl\langle \prod_{i\in {\cal I}} V_{\epsilon_i}(k_i)
\,V_\nu(q_{\cal I}, 0)\Bigr\rangle_W
\Bigl\langle \,V_\nu(-q_{\cal I})
\prod_{j\in {\cal J}} V_{\epsilon_j}(k_j)\Bigr\rangle_W
\nonumber
\ea
with 
$$
q = - \sum_{i\in {\cal I}} k_i.
$$
Here we used the notation
\be
\Bigl{\langle} \, \ldots
\Bigr\rangle_W = \int [dX] \, W[X] (\ldots).
\ee
The above form (\ref{newloop}) is quite suggestive of a string Ward 
identity. In the main text we try to make this correspondence more 
precise.\\

\medskip

\newsubsection{Appendix B: BRST derivation of the Ward identity}

A covariant definition of the scattering amplitude is given by
separating the anti-ghost fields, that provide the measure of 
integration on ${\cal M}_{D,n}$ from the vertex operators
\be
\label{covar}
\Bl \; \prod_{j} 
V_{\epsilon_j}(k_j) \; \Br_\Sigma
= \int_{\strut \!{\cal M}_{D,n}}\!\!\!\! \prod_i d\sss_i \
% d\sss_1 \ldots d\sss_{n-3} 
\Bigl\langle \prod_{i} V^{(-1)}_{\epsilon_i}(k_i,\sss_i)
\; \prod_{i} b_i \delta(\beta_i) \Bigr\rangle
\ee
with
\be
V^{(-1)}_{\epsilon}(k,\sss) = c(\sss) \delta(\gamma(\sss)) 
\epsilon^\mu \psi_\mu e^{ikX}
\ee
the photon emission vertex in the -1-picture. The condition of
conformal invariance is that the worldsheet stress-tensor $T(z)$ must
decouple from physical amplitudes
\be
\Bl \; T(z) \; \prod_{j} V_{\epsilon_j}(k_j) \; \Br_\Sigma =0
\ee
The insertion of the stress-energy tensor can be replaced by
\be
T(z) = \{Q_{brst}, b(z)\}.
\ee
In deriving the conformal Ward identity, we can make use of the 
BRST-invariance of the expectation value to perform a ``partial
integration'' and transpose the action of $Q_{brst}$ to the vertex 
operator insertions. Their transformation law reads
\be
\{Q_{brst}, V^{(-1)}_{\mu}(k,\sss)\} = 
\half k^2 \dot c(\sss) V^{(-1)}_{\mu}(k,\sss)
+ i k_\mu \dot \gamma(\sss)  V_T^{(-1)}(k,\sss)
\ee
with $V_T^{(-1)}$ the tachyon vertex (\ref{tachyon}) in the $-1$-picture.
After picture changing to make $V$ appropriate for integration
over the moduli $\sss_i$, the right-hand side becomes %\epsilon^\mu
\be
\half  (k^2 \delta_{\mu\nu}\!\! -\!
k_\mu k_\nu) \; \dot c \; V_{\nu}(k,\sss) 
+ \dddd (\, \dot c \;  k_\mu V_T^{(-1)}(k))
\ee
The total derivative term will drop out of (GSO projected) physical 
amplitudes. In addition, however, 
there is a contribution that arises because the 
BRST-charge acts as the exterior derivative on the moduli space:
the BRST-commutator with the anti-ghost insertions in (\ref{covar})
combines into a total derivative on ${\cal M}_{D,n}$. This total
derivative will 
integrate to to a non-zero result due to boundary contributions at 
$\partial{\cal M}_{D,n}$, which arise whenever
an intermediate photon state gets on-shell. A rather standard 
analysis shows that the resulting non-linear recursion relation, that
expresses the decoupling of the stress-energy tensor, takes the
form as given in (\ref{new}).

\renewcommand{\Large}{\large}

\end{document}